\documentclass[prb,twocolumn,showpacs,floatfix]{revtex4-1}
\usepackage{graphics,epsfig,amsfonts,amssymb,amsmath,soul}
\usepackage[T1]{fontenc}
\usepackage[utf8]{inputenc}
\usepackage{lmodern}
\usepackage[english]{babel}
\begin{document}
\selectlanguage{english}

\title{Bloch line dynamics within moving domain walls in 3D ferromagnets}

\author{Touko Herranen}
\email{touko.herranen@aalto.fi}
\author{Lasse Laurson}

\affiliation{COMP Centre of Excellence and Helsinki Institute of Physics,
Department of Applied Physics, Aalto University, P.O.Box 11100, 
FI-00076 Aalto, Espoo, Finland.}

\begin{abstract}
We study field-driven magnetic domain wall dynamics in garnet strips by 
large-scale three-dimensional micromagnetic simulations. The domain wall 
propagation velocity as a function of the applied field exhibits a low-field 
linear part terminated by a sudden velocity drop at a threshold field 
magnitude, related to the onset of excitations of internal degrees of
freedom of the domain wall magnetization. By considering a wide range of 
strip thicknesses from 30 nm to 1.89 $\mu$m, we find a non-monotonic 
thickness dependence of the threshold field for the onset of this instability, 
proceeding via nucleation and propagation of Bloch lines within the domain wall. 
We identify a critical strip thickness above which the velocity drop is due 
to nucleation of horizontal Bloch lines, while for thinner strips and depending on 
the boundary conditions employed, either generation of vertical Bloch lines, or 
close-to-uniform precession of the domain wall internal magnetization takes place. 
For strips of intermediate thicknesses, the vertical Bloch lines assume a 
deformed structure due to demagnetizing fields at the strip surfaces, 
breaking the symmetry between the top and bottom faces of the strip, and 
resulting in circulating Bloch line dynamics along the perimeter of the domain 
wall.
\end{abstract}
\maketitle

\section{Introduction}

Dynamics of driven domain walls (DWs) in ferromagnets of various confined geometries
such as wires, strips and thin films is an important problem both from the point-of-view
of technological applications, as well as due to strong fundamental physics interests.
One central feature of DW dynamics driven by applied magnetic fields \cite{beach2005, 
dourlat2008, metaxas2007, wb1974} or spin-polarized electric currents \cite{thiaville2005,
moore2008,tanigawa2008,parkin2008,fukami2008,thomas2006} is that the relation between the 
driving force (i.e., field or current density) and the resulting DW velocity tends to be 
non-monotonic: in addition to contributing to DW propagation, a sufficiently strong driving 
force may excite internal degrees of freedom of the domain wall, resulting in an abrupt drop
in the force-velocity curve of the DW. Within the one-dimensional (1D) model \cite{malozemoff1979},
these internal degrees of freedom are described by an angle corresponding to the 
orientation of the DW internal magnetization, which starts precessing above the Walker 
field or current density, resulting in an abrupt drop in the DW propagation velocity.

While the above description in terms of the 1D model should apply for narrow nanowires
and nanostrips, the situation is more complex when DWs in strips with a non-negligible 
width and/or thickness are considered. There, the excitation of the DW internal degrees
of freedom, taking place concurrently with a drop in the DW propagation velocity, may
be spatially non-uniform, and thus cannot be described by a single angular variable. In 
particular, in sufficiently wide thin ferromagnetic strips with perpendicular magnetic
anisotropy (PMA), the velocity drop takes place via nucleation and subsequent propagation
along the domain wall of vertical Bloch lines 
(VBLs) \cite{malozemoff1979,herranen2015}, or transition regions of different chiralities 
of the Bloch DW along its long axis. For thick enough strips or films with PMA, another 
type of excitation is expected to become prominent, namely the nucleation of 
horizontal Bloch lines (HBLs) \cite{malozemoff1979}; there, the 
DW chirality changes when moving along the DW in the thickness direction of the sample. 
These Bloch line structures, in particular in various garnet films, have been 
intensively studied already in the 1970s as they were at the time seen as potential 
building blocks of novel types of memory devices, the magnetic bubble 
memories \cite{bonyhard1973,nielsen1974,bobeck1975,bonyhard1976}.

Many of the related key studies of bubble materials such as garnet films where the 
presence of Bloch lines is essential for DW dynamics consist of theoretical work 
coupled with experimental observations \cite{slonczewski1973,thiaville1991,patek1994,
thiaville2001}. Due to recent advances in numerical 
techniques and the available computing power, it is now possible to perform full 
micromagnetic simulations of 3D samples with linear sizes reaching several microns, 
thus approaching the thickness range of typical garnet films studied in the past. In 
such simulations one may monitor the full 3D dynamics of the system, and thus 
obtain a more complete picture of the DW dynamics as compared to the typical 
experiments where one could observe only the surface of the relatively thick film. 

Thus, we perform here full 3D micromagnetic simulations of field-driven
DW dynamics in garnet strips with a wide range of strip thicknesses, considering as 
an example material the $\text{(GdTmPrBi)}_3 \text{(FeGa)}_5 \text{O}_{12}$ magnetic 
garnet \cite{thiaville2001}. Such materials may be grown epitaxially on a substrate, 
inducing a crystalline PMA \cite{malozemoff1979}, independent of the film/strip thickness, 
making it possible to systematically study the thickness dependence of the DW dynamics. 
For both periodic and open boundary conditions along the strip width, 
we find a low-field constant DW 
mobility regime, terminated at a sudden velocity drop at a threshold field magnitude. 
The threshold field and the corresponding (local) maximum of the DW propagation velocity 
exhibit a non-monotonic dependence on the sample thickness, with a peak of the maximum 
velocity occurring in films of a thickness related to the HBL width. We investigate 
in detail the related excitations of the DW internal structure, and find that for 
thin strips with open boundary conditions at the strip edges, the velocity drop is due to 
a VBL being nucleated from one of the strip edges, followed by its repeated propagation 
along the DW across the strip width. In contrast, the corresponding instability in 
thin samples with periodic boundary conditions 
proceeds 
via spatially close-to-uniform precession of the DW internal magnetization. In samples of 
intermediate thickness, the VBL structure is deformed due to the flux-closing 
tendency at the sample surfaces. This deformation results in interesting dynamics 
where the high spin rotation part of the deformed VBL repeatedly rotates around the strip 
along the edges and surfaces of the strip, thus breaking the symmetry between the top 
and bottom strip surfaces. For the thickest films considered (up to 1.89 $\mu$m), the 
velocity drop is related to a HBL being nucleated from one of the sample surfaces, which 
subsequently moves back and forth along the DW in the thickness direction of the 
sample. 

\begin{figure}[t!]
\includegraphics[width=\columnwidth]{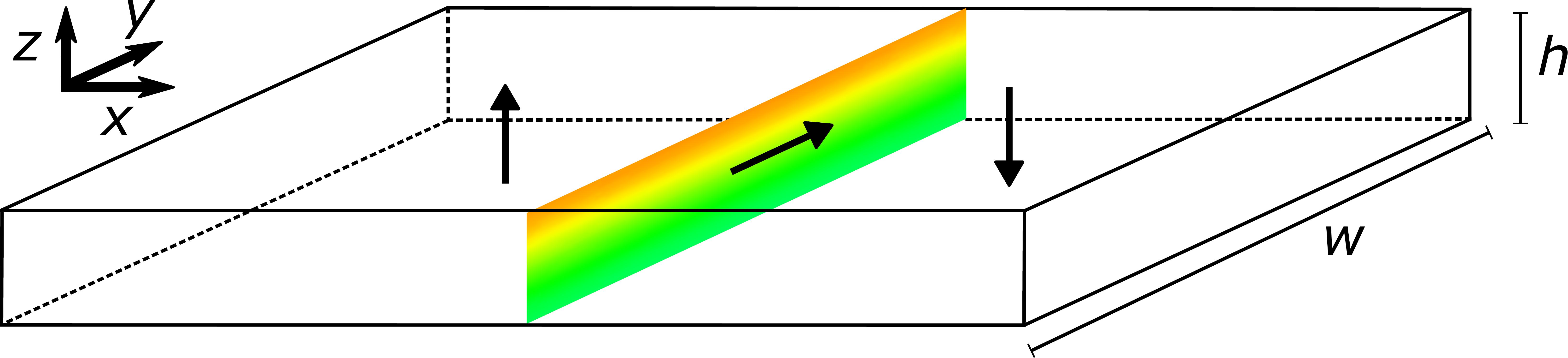} 
\caption{Relaxed initial magnetization configuration for $B_\text{ext}=0$ with 
a Bloch DW in the middle of a strip of thickness $h$ = 990 nm, separating the
$\pm z$ magnetized domains. 
The arrows indicate the magnetization direction. Due to the demagnetizing
fields acting on the DW and originating from the surface charges, the in-plane 
magnetization of the DW (visualized by the different colors) at the strip surfaces 
tilts from the $+y$ direction (light green) approximately by $37^\circ$ 
towards $\pm x$ direction (orange and darker green, respectively) for this particular 
$h$-value, while the pure Bloch wall structure (DW magnetization along $+y$) is 
maintained for $z=h/2$.}
\label{fig:schema}
\end{figure}

The paper is organized as follows: In Section \ref{sec:methods}, 
we describe the details of our micromagnetic simulations, including material parameters 
used, and the geometry of the samples considered. In Section \ref{sec:results}, we 
study the DW velocity $v$ as a function of the driving field strength $B_\text{ext}$, 
sample thickness $h$ for the different boundary conditions, focusing in particular
on the different excitation modes responsible for the velocity drop in the 
$v(B_\text{ext})$ curves. In Section \ref{sec:discussion}, we finish the paper by 
presenting our conclusions.

\section{Methods}
\label{sec:methods}
The micromagnetic simulations are performed using the GPU-accelerated micromagnetic 
code MuMax3 \cite{mumax_git,mumax2011,mumax2014}, which solves numerically  
the Landau-Lifshitz-Gilbert equation \cite{gilbert2004,brown1963},
\begin{equation}
\label{eq:llg}
\partial {\bf m}/\partial t =
\gamma {\bf H}_\text{eff} \times {\bf m} + \alpha {\bf m} \times
\partial {\bf m}/\partial t,
\end{equation}
where ${\bf m} = {\bf M}/M_\text{S}$ is the magnetization, $\gamma$ the 
gyromagnetic ratio, $\alpha$ the Gilbert damping parameter, and 
${\bf H}_\text{eff}$ the effective field, with contributions due to exchange, 
anisotropy, Zeeman, and demagnetizing energies. 

We simulate DW dynamics in garnet strips with a wide range of thicknesses. 
As a test material we choose to consider $\text{(GdTmPrBi)}_3 \text{(FeGa)}_5 \text{O}_{12}$ 
magnetic garnet with saturation magnetization $M_\text{S}$ = 8992 A/m, uniaxial 
out-of-plane anisotropy constant $K_\text{u}$ = $602.5$ $\text{J/m}^3$, exchange 
constant $A$ = $2.2 \cdot 10^{-12}$ J/m, and damping parameter 
$\alpha$ = $0.15$ \cite{thiaville2001}. The quality factor for this material is 
$Q = K_\text{u} / K_\text{d} =$ 11.9, with the stray field energy constant given by 
$K_\text{d} = {M_\text{S}}^2 \mu_0/2 \approx 50.8 $ $\text{J/m}^3$, where $\mu_0$ is vacuum permeability. The Bloch wall width 
parameter $\Delta = \sqrt{A/K_\text{u}} \approx $ 60 nm.

In the simulations, we fix the sample width to $w$ = 3.84 $\mu$m, and use a 
moving simulation window of length 30.72 $\mu$m centered around the DW; during 
the simulation the simulation window is shifted such that the average DW $x$ 
position is always kept within one discretization cell from the middle of the 
sample in the $x$ direction.
The thickness of the sample is varied between $h$ = 30 nm and $h$ = 1.89 $\mu$m. We 
use cubic discretization cells with a side length of $\Delta_\text{cell}  = $ 30 nm 
$\approx \Delta/2$ \cite{rave1998}.

The magnetization is initialized into two domains, with magnetization along $+z$ 
and $-z$ directions, respectively, with a $+y$ magnetized Bloch wall separating 
the domains in the middle of the sample. The system is then let relax to its 
equilibrium configuration. A relaxed 
micromagnetic configuration for $h$ = 990 nm is presented schematically in 
Fig. \ref{fig:schema}. The domains generate magnetic surface charges, which in turn 
create in-plane demagnetizing fields acting on the DW magnetization at the sample
surfaces. In a static ($B_\text{ext} = 0$) isolated DW, the demagnetizing field 
tends to tilt the DW magnetization towards a N\'eel structure close to the top and 
bottom surfaces, in order to close the flux \cite{schlomann1974,malozemoff1979}.
The center of the wall remains in a pure Bloch wall structure. Theoretical calculations by 
Slonczewski show that the demagnetizing field component perpendicular to the wall in an 
isolated DW is $H_S(z) = 4 M_S \ln \left[ z/\left( h - z \right) \right]$ in the limit 
$\Delta / h = 0$
\cite{schlomann1974}, with the sample surfaces at $z = \lbrace 0,h \rbrace$.
Using this expression, one may define the so-called critical points where $|H_S(z)| = 8 M_S$,
located at $z_{a}(h)=h/(1+e^2)$ and $z_{b}(h)=he^2/(1+e^2)$, respectively \cite{schlomann1974}.
These are understood as points where HBLs may be nucleated. 
The twist of the DW magnetization due to demagnetizing fields is suppressed by 
the exchange stiffness for thin samples \cite{hubertschafer}.

In what follows we present an extensive micromagnetic study of the DW dynamics 
induced by driving fields $B_\text{ext}$ of different magnitudes along the $+z$ 
direction, by varying the boundary conditions (open vs periodic along the strip width), 
and the sample thickness $h$. More specifically, we address the question of how the 
dynamics of the internal magnetization of the DW affects the DW propagation velocity, 
focusing in particular on the excitations responsible for the sudden velocity drops 
in the $v(B_\text{ext})$ curves.

\begin{figure*}[t!]
\centering
\includegraphics[width=1.7\columnwidth]{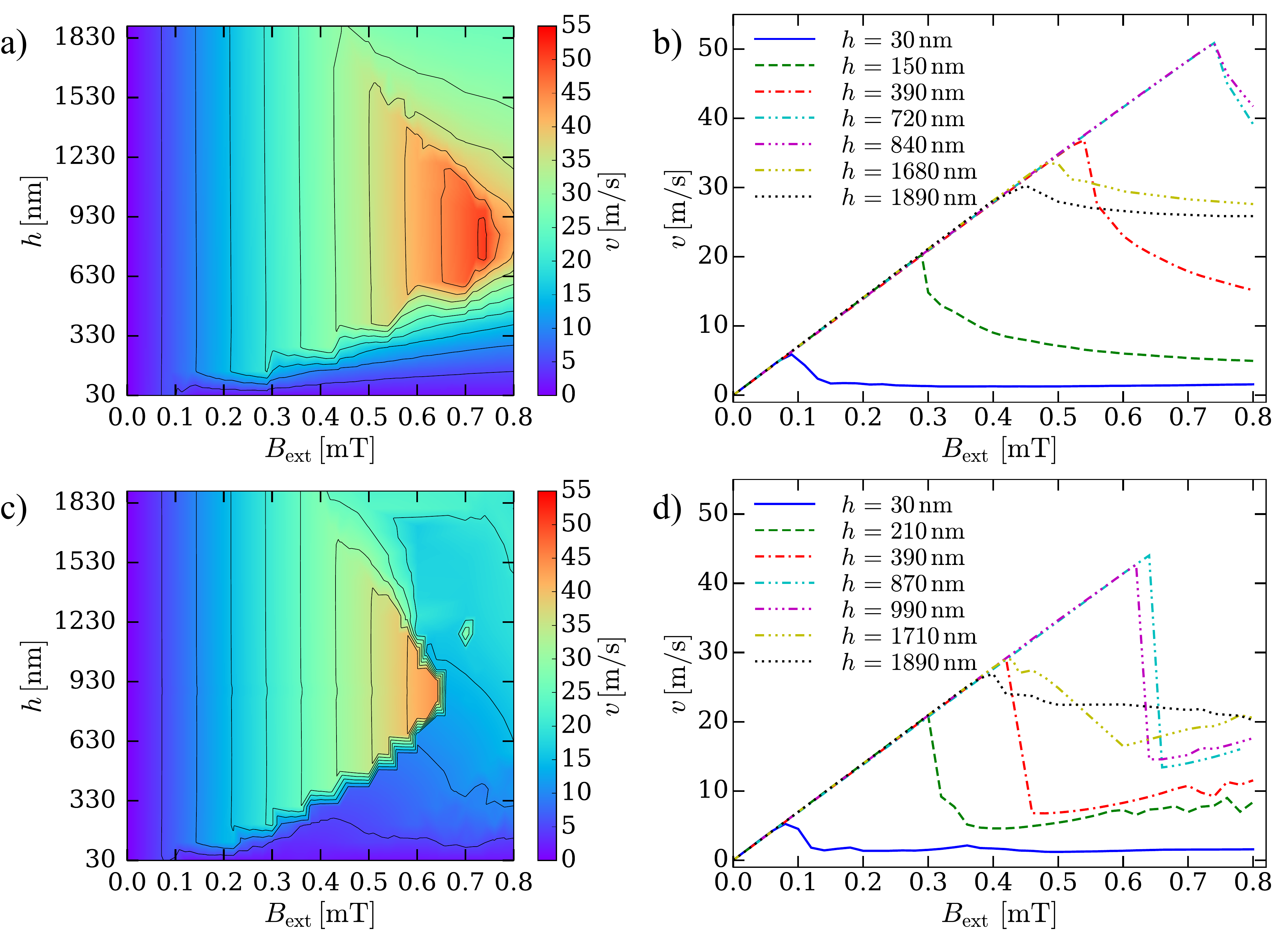} 
\caption{Average field-driven DW propagation velocities $v$ as a function of $B_\text{ext}$
and $h$ for PBCs [a) and b)] and OBCs [c) and d)] in the $y$ direction. In each case,
the velocities are calculated from the slope of a linear fit to the DW position $x(t)$,
excluding the initial transient. For $B_\text{ext}>B_\text{ext}^\text{max}$,
the transient time is estimated to be the time when the initially positive $y$ component
of the DW magnetization first changes sign (indicating the onset of periodic precessional
dynamics), while for smaller fields a fixed 100 ns transient time is removed from the
start of each simulation run.}
\label{fig:v_all}
\end{figure*}

\section{Results}
\label{sec:results}

To understand the details of field-driven domain wall dynamics in garnet
strips of different thicknesses, we consider here separately two different boundary
conditions along the $y$ direction (see Fig. \ref{fig:schema}), i.e., periodic
boundary conditions (PBCs) mimicking an infinitely long DW, and open boundary
conditions (OBCs) where the effects due to strip edges are important. In both
cases, we measure the relation between the DW propagation velocity $v$ and the
applied field $B_\text{ext}$ along the $+z$ direction, i.e., $v(B_\text{ext})$,
and focus on the excitations of the DW internal magnetization responsible for
the sudden drop of $v(B_\text{ext})$ at a threshold field strength, and how their
nature depends on the boundary conditions and the sample thickness $h$. We consider
fields up to $B_\text{ext} = 0.8$ mT, enough to observe the velocity breakdown
in all the systems considered.

\subsection{Periodic boundary conditions}

We start by considering PBCs in the $y$ direction. The resulting DW propagation
velocities as a function of $B_\text{ext}$ and $h$ are presented in Figs. 
\ref{fig:v_all} a) and b). For $B_\text{ext}$ below the $h$-dependent threshold
field (also referred to as the critical field in what follows) for the onset of 
the instability resulting in a velocity drop, $v(B_\text{ext})$
is linear in $B_\text{ext}$. Within this linear regime, the DW magnetization remains
constant during DW motion after an initial transient, and the DW mobility 
$\mu = \text{d}v/\text{d}B_\text{ext}$ is well-described by the well-known theoretical result of 
$\mu_\text{steady} = \gamma \Delta / \alpha \approx 71 \; \mathrm{ms^{-1}mT^{-1}}$ 
for all $h$ \cite{malozemoff1979}. 

\begin{figure}
\includegraphics[width=0.9\columnwidth]{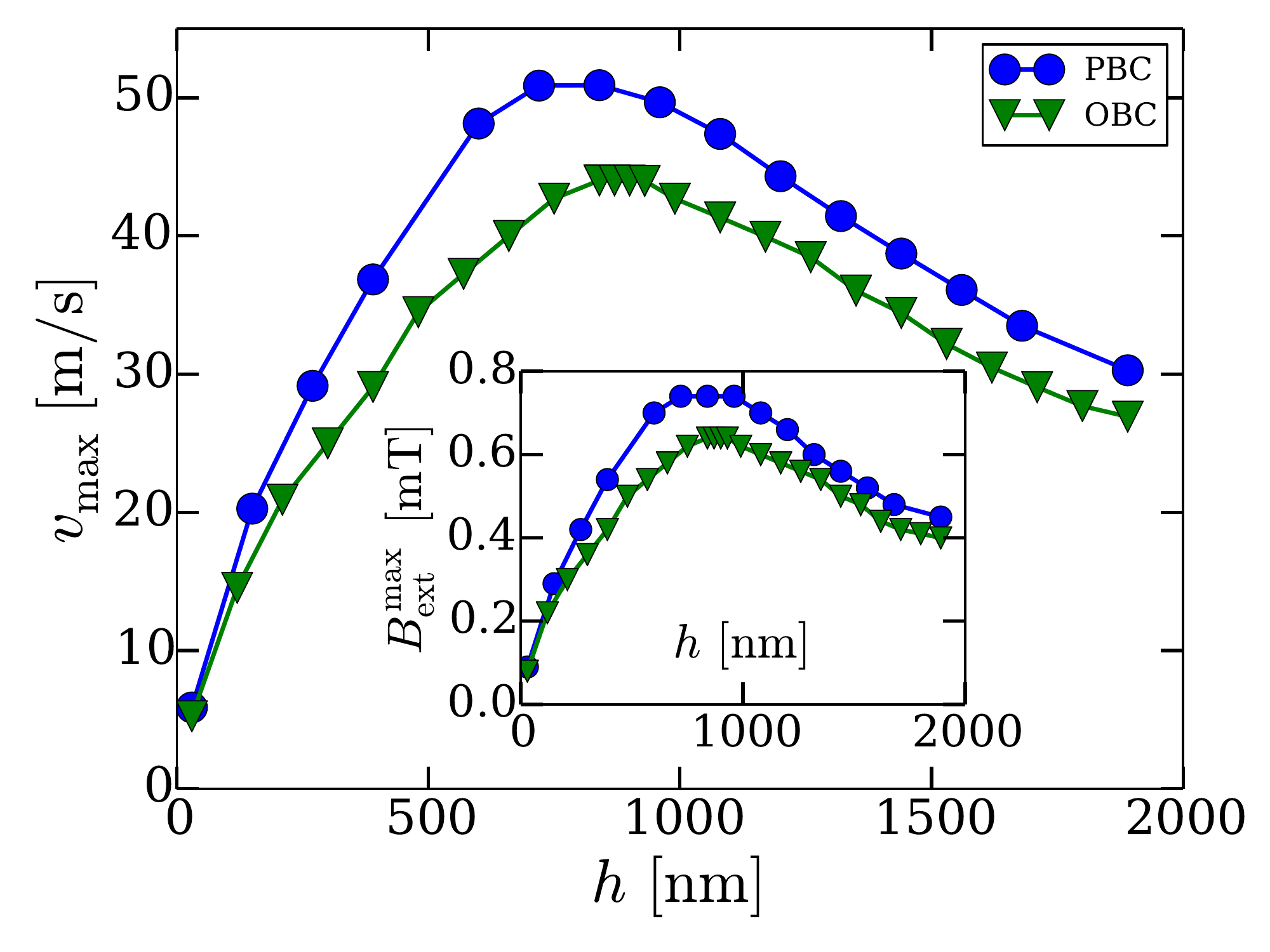} 
\caption{Maximum of the average field-driven DW propagation velocity $v_\text{max}$ 
before the velocity drop as a function of the sample thickness $h$. The inset depicts 
the $h$-dependence of the corresponding applied field value $B_\text{ext}^\text{max}$. 
OBCs in the $y$ direction give rise to lower values of $v_\text{max}$ (and 
$B_\text{ext}^\text{max}$), given the additional excitation modes available due to 
open boundaries at the strip edges.}
\label{fig:vmax}
\end{figure}

The $h$-dependent peak DW propagation velocities $v_\text{max}$, extracted from data 
in Fig. \ref{fig:v_all} considering PBCs along $y$, are presented in Fig. \ref{fig:vmax}
(blue circles).
The corresponding threshold fields for the onset of the instability, i.e., the 
fields $B_\text{ext}^\text{max}$ at which the (local) maximum velocity $v_\text{max}$ occurs, 
are reported in the inset of 
Fig. \ref{fig:vmax} as a function of $h$. Both $v_\text{max}$ and $B_\text{ext}^\text{max}$
display an intriguing non-monotonic dependence on $h$. To understand this, we need
to consider the different excitation modes of the DW internal magnetization responsible
for the onset of the velocity drop, and how these depend on $h$.

Thus, we consider next the dynamics of the DW internal magnetization just above the
$h$-dependent critical field $B_\text{ext}^\text{max}$ corresponding to the local maximum 
of the DW propagation velocity. In thin samples ($h \leq$  150 nm), we observe spatially 
close-to-uniform rotation of the DW magnetization when fields just above the critical 
field are applied (see Supplemental Material, Movie 1 \cite{SM}), 
similarly to the Walker breakdown dynamics observed in nanowire 
geometries, and as predicted by the 1D model. As is visible in Movie 1, the spatially 
uniform  DW magnetization 
rotation does not occur at a constant rate. The rotation is slowed down when the DW 
magnetization $m_\text{DW}$ assumes a tilted (due to the finite $B_\text{ext}$) Bloch wall 
configuration. After $B_\text{ext}$ has slowly rotated $m_\text{DW}$ from the tilded Bloch 
towards a N\'eel configuration, $m_\text{DW}$ rotates abruptly to reach the opposite tilted 
Bloch wall configuration, where the magnetization rotation slows down again.

\begin{figure}
\includegraphics[width=\columnwidth]{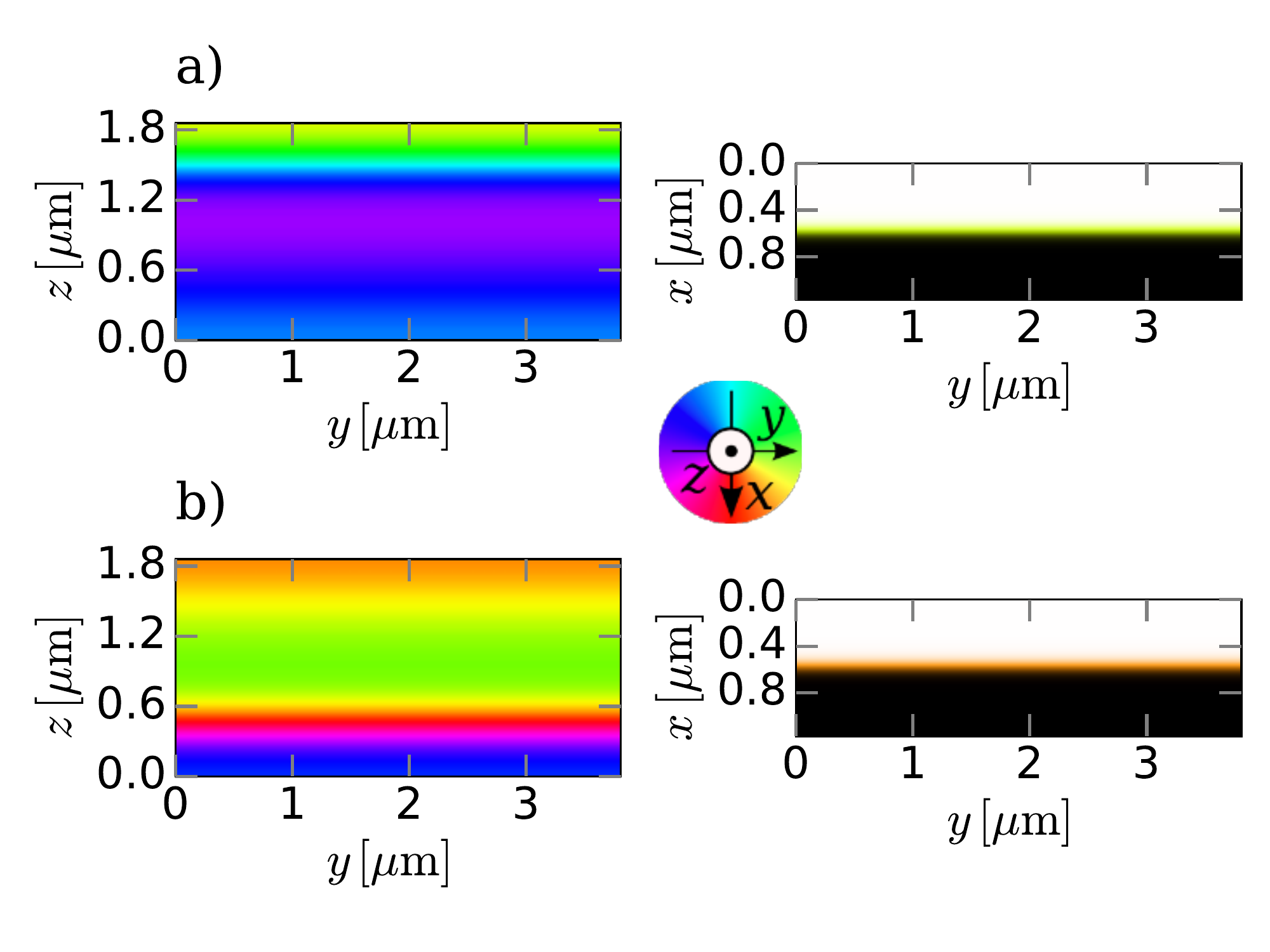} 
\caption{A HBL subject to an applied field of $B_\text{ext} = 0.6$ $ \text{mT} $ traveling 
along the DW in the $z$ direction in a strip of thickness 
$h = $ 1.89 $\mu$m with PBCs in the $y$ direction. 
Left panels show snapshots of the magnetization configuration within the plane of
the DW (defined as $m_z=0$), while the right panels show the corresponding magnetization 
that could be observed on the top surface of the sample. In a), the HBL 
(with the mid-point magnetization $m_x^\text{BL} = -1$, cyan color) is moving 
upwards (in the $+z$ direction) along the DW. Shortly after the snapshot shown in a), 
it punches through the top surface. Soon after that a new HBL with opposite magnetization 
($m_x^\text{BL} = +1$, red color) is nucleated and it travels to the opposite direction, 
i.e., in the $-z$ direction. The snapshot in b) shows the HBL just before it punches
through the bottom surface, after which the process repeats. The left panels of a) and
b) show the magnetization in the DW plane (defined as $m_z=0$), while the right panels 
show the corresponding strip magnetization in the top surface of the strip. 
The color wheel shows the mapping between colors and the direction of the in-plane 
magnetization, and white and black correspond to $m_z=1$ and -1, respectively.}
\label{fig:array_hbl}
\end{figure}

When $h$ is increased 
towards ``intermediate'' thicknesses (150 nm $< h <$ 720 nm), we start to see some 
internal structure developing in the DW magnetization during the magnetization 
precession process. More specifically, we observe a ``partial'' HBL (meaning 
that it does not exhibit a full $\pi$ rotation of the magnetization) repeatedly 
nucleating from one of the sample surfaces and subsequently traveling along the 
DW in the $z$ direction, leading to repeated switching of the DW internal magnetization. 
For samples with even larger thicknesses, a ``full'' $\pi$ HBL structure is 
observed to nucleate and propagate along the DW in the thickness direction of the 
sample, see Fig. \ref{fig:array_hbl}, as well as Supplemental Material, 
Movie 2 \cite{SM}. There, the HBL travels along the DW in 
the $\pm z$ directions, and punches through at the top or bottom surface, with another 
HBL nucleated shortly after each punch-through. The newly nucleated HBL has opposite 
in-plane magnetization and travel direction to the previous one.

The question is now how the above observations about the internal magnetization dynamics
of the DW may be used to understand the non-monotonic $h$-dependence of $v_\text{max}$
(and consequently, of $B_\text{ext}^\text{max}$). 
Theoretical calculations by Slonczewski and Malozemoff \cite{slonczewski1973,malozemoff1979} 
suggest that for samples with thicknesses above the so-called Bloch line limit, 
i.e., for $h \gg \Lambda$, where $\Lambda = \sqrt{A/K_\text{d}} 
\approx $ 208 nm is the Bloch line width parameter, the maximum velocity obeys $v_\text{max}(h) \sim 1/h$; 
similar behavior with $v_\text{max}$ decreasing with $h$ can be seen in our results 
for the largest thicknesses considered. 

On the other hand, for thin strips where the DW internal magnetization exhibits 
spatially uniform precession, one may apply the theory of Mougin {\it et 
al.} \cite{mougin2007}, stating that in confined geometries the Walker field 
assumes a modified form $H_\text{W} =  2\pi\alpha M_\text{S} |N_x-N_y|$, where 
$N_x$ and $N_y$ are the demagnetizing factors along $x$ and $y$, respectively. 
For the present case of PBCs along $y$, the DW length is effectively infinite, 
and thus $N_y=0$. Using the elliptic approximation for $N_x$ \cite{mougin2007}, 
one obtains $H_\text{W} = 2\pi \alpha M_\text{S} h/(h+\Delta)$, and $v_\text{max} = 
(2\pi \gamma \Delta M_\text{S} h)/(h+\Delta)$. Both of these increase with $h$ for small $h$, in qualitative 
agreement with our results.

Thus, it appears that our results for the PBC case interpolate between a 
small-$h$ regime where the onset of the uniform precession of the DW internal
magnetization is controlled by the thickness dependent demagnetizing factor
$N_x$ of the DW (leading to a $v_\text{max}$ increasing with $h$), and a 
large-$h$ regime controlled by the nucleation and subsequent dynamics of a HBL,
leading to a decreasing $v_\text{max}$ with $h$ (asymptotically of the form of 
$v_\text{max} \sim 1/h$ \cite{slonczewski1973,malozemoff1979}).
These two regimes are separated by a maximum of $v_\text{max}$, which for
our system occurs at $h \approx$ 840 nm. For this thickness, $h$ roughly
coincides with $z_a(h) + [h-z_b(h)] + \pi \Lambda \approx $ 850 nm where 
$z_a(h)$ and $z_b(h)$ are the critical points, and $\pi\Lambda$ is the natural BL width \cite{malozemoff1979}. 
Thus, the HBL width together with the additional thickness due to the two critical 
points appears to set a characteristic thickness corresponding to the maximum of 
the $v_\text{max}(h)$ curve.
In other words, as soon as the sample thickness $h$ is great enough to accommodate
a full $\pi$ HBL as well as the critical points to nucleate HBLs, the energy
barrier to nucleate HBLs becomes lower and the maximum achievable stable velocity
$v_\text{max}$ starts to decrease with increasing $h$.

\begin{figure}
\includegraphics[width=\columnwidth]{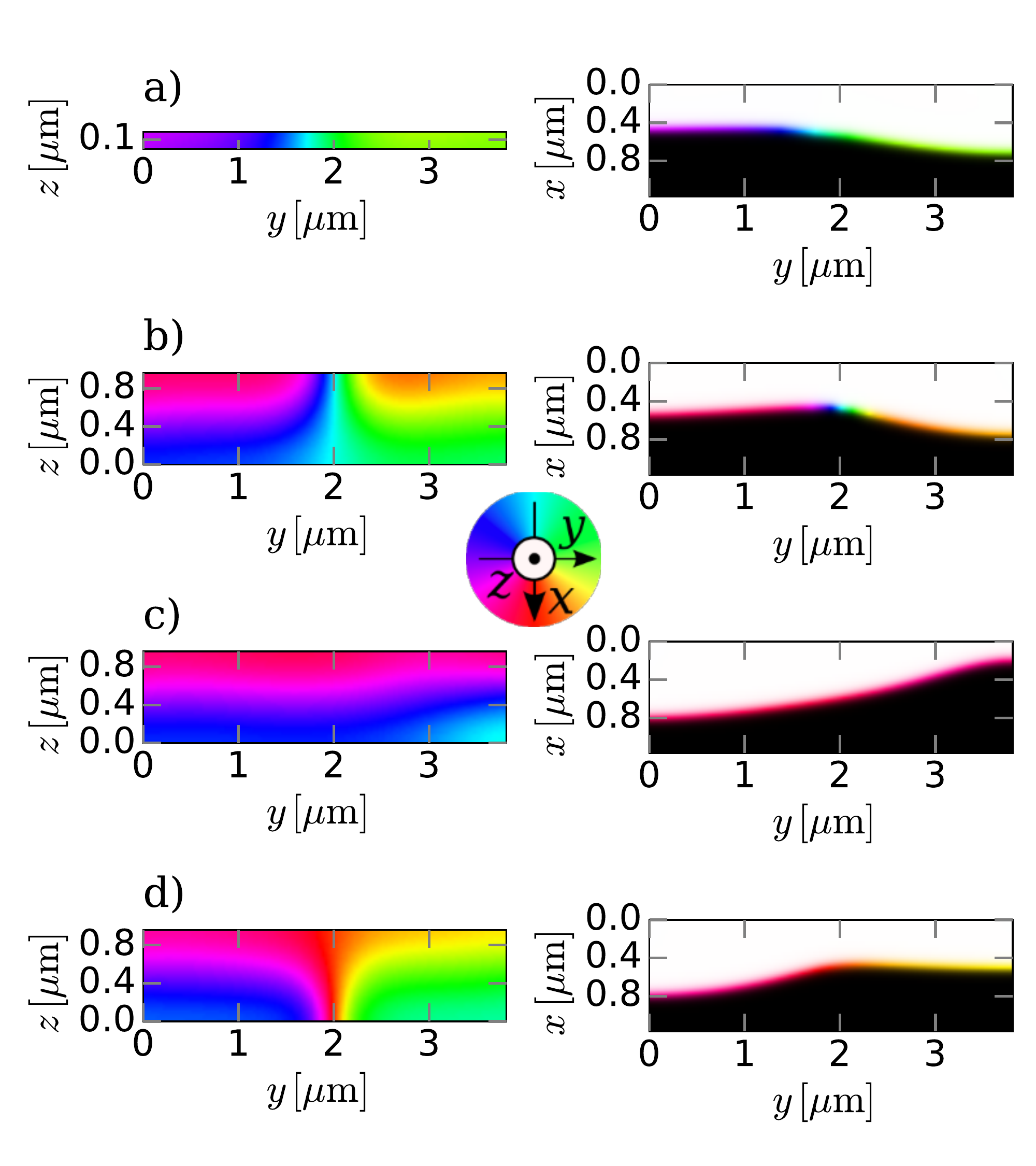} 
\caption{VBL dynamics in systems with OBCs in the $y$-direction. 
Left panels show snapshots of the magnetization configuration within the plane of
the DW (defined as $m_z=0$), while the right panels show the corresponding magnetization 
that could be observed on the top surface of the sample. a)
shows a snapshot from a strip with thickness of $h$ = 210 nm, $B_\text{ext}= 0.32 $ $\text{mT}$,
with a VBL traveling in the $+y$ direction. b), c) and d) display three snapshots from
a system with $h$ = 990 nm and $B_\text{ext}=0.64$ mT, with a deformed VBL with the mid-point
VBL magnetization along $-x$ (cyan) propagating in the $+y$ direction in b), followed 
by a transient HBL-like structure in c). After that, as shown in d), a deformed VBL 
with the mid-point magnetization along $+x$ moves in the $-y$ direction. Notice how
the top view of the sample in d) hardly contains any direct information about the
presence of a VBL within the DW, apart from the retarded part at the location of the VBL.
The color wheel shows the mapping between colors and the direction of the in-plane
magnetization, and white and black correspond to $m_z=1$ and -1, respectively.}
\label{fig:array}
\end{figure}

\subsection{Open boundary conditions}

Next, we proceed to consider the DW dynamics in a system with OBCs,
to understand the effect of the presence of strip edges on the onset of 
the excitations of internal degrees of freedom of the DW internal magnetization.
The resulting DW propagation velocities $v(B_\text{ext},h)$ are presented 
in Figs. \ref{fig:v_all} c) and d). The same linear part for small fields as in the 
PBC case can be observed, but the velocity drop at which the linear part terminates 
is more abrupt and pronounced for OBCs. Also, as can be seen in Fig. \ref{fig:vmax} 
(green triangles), the system with OBCs reaches the maximum velocity for each $h$ 
at a lower field than the corresponding PBC system, and consequently the value of 
$v_\text{max}$ is also lower. This difference between PBCs and OBCs
is expected as the OBCs provide additional possibilities to nucleate excitations
such as vertical Bloch lines from the strip edges, not present in the system with PBCs.
This observation may also be rationalized within the theory of Mougin {\it et al.} 
\cite{mougin2007} discussed above, by noticing that for OBCs $N_y \neq 0$, leading 
to a smaller value of $H_W$ (and thus of $v_\text{max}$) as compared to the PBC 
case where $N_y=0$. Moreover, by comparing the $v(B_\text{ext},h)$ data for OBCs 
and PBCs [Figs. \ref{fig:v_all} b) and d), as well as Fig. \ref{fig:vmax}], it 
appears that the nature of the DW dynamics becomes increasingly similar in the two 
cases as $h$ increases (i.e., the boundary conditions appear to become asymptotically
irrelevant for large $h$). In particular, for the thickest system considered with 
$h = $ 1.89 $\mu$m, $v_\text{max}$ and $B_\text{ext}^\text{max}$ have quite 
similar values for PBCs and OBCs (Fig. \ref{fig:vmax}). This is understandable
as based on our observations for PBCs discussed above, in thick enough samples 
the breakdown dynamics is expected to be dominated by nucleation of HBLs at the 
strip surfaces, which should not be sensitive to the boundary conditions at the 
strip edges.

We then proceed to consider the $h$-dependent excitations of the DW internal 
magnetization responsible for the velocity drop visible in Fig. \ref{fig:v_all} d), 
considering again the dynamics taking place for the smallest field value above 
$B_\text{ext}^\text{max}$ for each $h$. 
In thin systems ($h <  210$ nm) we observe a VBL repeatedly nucleating at one of the 
strip edges, and subsequently traveling towards the other edge [see Fig. 
\ref{fig:array} a) for a snapshot of a traveling tail-to-tail VBL in a system with 
$h$ = 210 nm; see also Supplemental Material, Movie 3]. This dynamics is similar 
to the dynamics found recently in simulations 
of thin CoPtCr samples with OBCs \cite{herranen2015}. For somewhat thicker strips ($210$ 
nm $< h <  390$ nm) we observe a VBL nucleating from one of the strip edges, 
and subsequently traveling back and forth between the strip edges. 
Moreover, in this thickness range, the VBLs start to become slightly deformed due to 
the flux-closing tendency at the strip surfaces: as discussed before, the surface 
charges due to the $\pm z$ magnetized domains result in demagnetizing fields
acting on the DW magnetization and pointing along positive and negative $x$ 
directions at the top and bottom surfaces of the strip, respectively. These fields
tend to tilt the magnetization surrounding the midpoint of the VBL away from the
otherwise preferred $\pm y$ directions (i.e., from the Bloch wall structure), 
leading to a deformed VBL structure breaking the symmetry between the top and 
bottom surfaces of the strip.

Upon further increasing $h$, this VBL deformation becomes more pronounced, and their 
dynamics change drastically. 
During the propagation across the strip width along the DW, on one strip surface (face) 
the VBL makes almost a full $2\pi$ rotation, while on the opposite face the 
magnetization rotates roughly by $\pi/2$ only. After reaching the other strip edge, the
system exhibits a short-lived transient partial HBL structure (in that the magnetization 
rotation is less than $\pi$ due to the limited strip thickness), followed by the 
propagation of a deformed VBL towards the other strip edge. Figs. \ref{fig:array} b), 
c) and d) (and Supplemental Material, Movie 4 \cite{SM}) illustrate this in the case 
of a tail-to-tail VBL within a $h = $ 990 nm sample; 
here, the head-to-head vs tail-to-tail nature of the deformed VBL may be defined by 
considering the magnetization profile along the DW in the middle layer of the strip, 
where the magnetostatic effects due to the surface charges are negligible. 
Fig. \ref{fig:array} b) shows a snapshot of a deformed VBL traveling in the
positive $y$ direction. The view on the top face of the strip (shown on the right)
illustrates the close to $2\pi$ rotation of the DW magnetization one would observe
by just looking at the sample surface. Fig. \ref{fig:array} c) presents a snapshot
of the partial HBL structure, which then transforms into the deformed VBL traveling
in the negative $y$ direction as shown in Fig. \ref{fig:array} d). Notice the small
magnetization rotation ($\sim \pi/2$) observable on the top face of the strip 
[Fig. \ref{fig:array} d), right panel]. The next step is the formation of another
HBL-like structure (not shown), after which the process repeats itself.
The circulation direction [clockwise as in Figs. \ref{fig:array} b), c) and d),
or counter-clockwise] is determined by the topology of the VBL: the large spin rotation part
of deformed tail-to-tail 
VBLs exhibit clockwise rotation, while head-to-head VBLs circulate counter-clockwise.
In general, the propagation direction of the various Bloch line structures is determined
by the direction in which the applied field tends to rotate the mid-point magnetization
of the Bloch line. Dynamics of these deformed VBL structures have been previously 
studied for fixed film thicknesses \cite{redjdal1996,bagneres1992,thiaville2001}, but 
to our knowledge the circulating VBL dynamics described above have not been reported 
before.

We note that when the part of the deformed VBL with more spin rotation is traveling 
along the bottom face of the strip [Fig. \ref{fig:array} d)], 
an experimental measurement of magnetization of the top surface of the sample would 
show no clear VBL structure but only a retarded  
part of the DW with a $\sim\pi/2$ rotation of the DW internal 
magnetization along the DW. 
On the bottom surface the magnetization rotates again by 
almost $2\pi$. The precise amount of spin rotation across the deformed VBL varies 
somewhat during its propagation, and depends also on $h$. Observation of a retarded 
region within the DW may thus provide an indirect way to experimentally detect 
VBLs within moving DWs in thick samples in situations where the part of the VBL with 
more spin rotation is hiding on the bottom surface of the strip. 

In strips with thicknesses $h > 1.71$ $\mu$m, the internal DW magnetization dynamics 
becomes similar to that in the corresponding systems with PBCs, i.e., a HBL is 
nucleated on one of the strip surfaces (instead of a VBL at one of the strip edges
in thinner strips). However, the punch-through mechanism of the HBL is 
different. As a HBL approaches a strip surface, both ends of the HBL (located
close to the corners of the strip) turn 
into structures reminiscent of the large spin rotation parts of the deformed VBLs 
discussed above, which then move 
towards each other along the surface, and eventually annihilate each other. Then,
another HBL is nucleated at this strip surface, and subsequently propagates towards
the other surface of the strip, and the process is repeated. This is illustrated
by Supplemental Material, Movie 5 \cite{SM}.

$h = $ 1.26 $\mu$m to $h = $ 1.71 $\mu$m is a transition thickness region, where 
both of the aforementioned breakdown dynamics are seen in the same system. 
Within this thickness range the dynamics begin with a few HBL punch-through events, 
as described above. After the punch-through events the structure changes into a deformed 
VBL, circulating around the DW similarly as described above for ``intermediate'' strip
thicknesses. The DW velocity $v(t)$ has two phases as well corresponding to these two
different kinds of BL dynamics: during HBL propagation the DW moves forward with a 
constant velocity, but during the punch-through, the DW stops and travels momentarily 
(for a time period of roughly $\sim$1 ns) backwards, and then proceeds forwards after 
nucleating a new HBL. In the VBL phase the domain wall is constantly moving forward 
with a slightly oscillating velocity, which on the average is lower than in the 
HBL phase. For instance, for $h = 1.71$ $\mu$m and $B_\text{ext} = $ 0.7 mT, the 
average DW velocity in the HBL phase is $\sim$5 m/s higher than in the VBL phase.
In Fig. \ref{fig:v_all}, we always report the steady state velocity after any
initial transients.
 
Finally, we should note that we found an anomaly at $h = 1.17$ $\mu$m and 
$B_\text{ext} = $ 0.7 mT visible in Fig. \ref{fig:v_all} c) as a small island of 
relatively fast DW velocity as compared to the background. Inspecting the magnetization
dynamics, we found that in this particular point it is similar to what is seen in 
the thickest systems with purely HBL-based BL dynamics with OBCs, while the surrounding
region exhibits mostly VBL dynamics. This could be, e.g., due to the shortness of the
simulation time, such that the system would not reach the true steady state
dynamics within the simulation time. We have verified that 
$B_\text{ext} = $ 0.7 mT $\pm \epsilon$, where $\epsilon = 0.001 $ mT, results in 
the same behavior.

\section{Discussion and Conclusions}
\label{sec:discussion}
Thus, we have shown that the maximum stable field-driven DW propagation velocity $v_\text{max}(h)$ 
in garnet strips exhibits a non-monotonic dependence on the strip thickness $h$.
We identify a characteristic sample thickness resulting in the maximum of $v_\text{max}(h)$
to be given roughly by the Bloch line width $\pi \Lambda$ of the material, together 
with the extra thickness needed to accommodate the critical points next to both strip 
surfaces, and describe the types of excitations (uniform magnetization rotation, nucleation
of VBLs of different types, as well as HBLs) responsible for the velocity drop for 
thicknesses above and below the characteristic one, considering both periodic and open 
boundary conditions at the strip edges. We were able to qualitatively account for the 
observations by comparing our results with previous theoretical analysis. For thin 
strips, the increasing trend of $v_\text{max}(h)$ with $h$ may be understood in terms 
of the work of Mougin \textit{et al}. \cite{mougin2007}, while in the limit of thick 
samples our results appear to approach those of Slonczewski, obtained for thicknesses 
well beyond the HBL limit, $h\gg \Lambda$ \cite{slonczewski1973}, leading to a 
decreasing $v_\text{max}(h)$ with $h$.  

Out of the excitation modes resulting in the drop of the DW propagation velocity, the 
circulating motion of the high spin rotation part of a deformed (due to demagnetizing 
fields originating from surface charges) VBL along the perimeter of the DW in strips 
of intermediate thickness with OBCs has - as far as we know - not been reported before. 
Experimental verification of such a mode would be an interesting avenue for future work; 
in the present paper, we provide some guidelines on how to achieve this.

Finally, we point out that extending our work to the case of 3D samples with 
quenched disorder \cite{leliaert2014numerical} 
interacting with the DW as well as with the various BL structures within the 
DW \cite{herranen2015} would be 
another important direction for forthcoming studies. In particular, it would allow to 
address the question of the possible relevance of the DW internal structure, including 
VBLs and HBLs, on the nature of the jerky, avalanche-like DW motion observed in the 
context of the Barkhausen effect \cite{zapperi2006barkhausen}. 
Most often such dynamics have been modeled by describing the DWs as elastic 
interfaces in random media, with the details of the interaction kernel depending
on whether the long-range dipolar interactions are thought to be relevant or not 
\cite{zapperi1998dynamics,laurson2014universality}. However, in such models the 
dynamical internal structure of the DW is neglected, and it is a pertinent question 
if such an approximation is valid in all cases.\\

\begin{acknowledgments}
This work has been supported by the Academy of Finland through its Centres
of Excellence Programme (2012-2017) under project no. 251748, and an Academy
Research Fellowship (LL, project no. 268302). We acknowledge the computational 
resources provided by the Aalto University School of Science ``Science-IT'' 
project, as well as those provided by CSC (Finland).
\end{acknowledgments}

\bibliography{bibl}

\end{document}